\documentclass{cmmse2014}
\usepackage{amssymb}
\usepackage[pdftex]{graphicx}
\usepackage{epstopdf}
\usepackage{color}
\usepackage{amsfonts}
\usepackage{amssymb}
\usepackage{caption}
\usepackage{graphicx}
\usepackage{float}
\usepackage{multirow}
\usepackage[utf8]{inputenc}
\usepackage{amsmath}
\usepackage{amssymb}

\usepackage[]{algorithm2e}
\usepackage{tikz}
\usetikzlibrary{automata,arrows,calc,positioning}
\usepackage{tabularx}
\newcolumntype{C}{>{\centering\arraybackslash}X}
\usepackage{setspace}
\usepackage{booktabs}
\newcolumntype{L}{>$l<$}
\usepackage{enumitem}
\usepackage{subfig}
\usepackage{algpseudocode}

\begin{document}



\newcommand{\Eqref}[1]{(\ref{#1})}
\newcommand\thefont{\expandafter\string\the\font}

\newcommand*{\underuparrow}[1]{\ensuremath{\underset{\uparrow}{#1}}}
\captionsetup{width=0.8\textwidth}


\markboth
  {} 
  {}


\title{Large Scale Automated Forecasting for Monitoring Network Safety and Security}



\author{Roi Naveiro}{roi.naveiro@icmat.es}{1}

\author{Sim\'on Rodr\'iguez}{simon.rodriguez@icmat.es}{1}

\author{David R\'ios Insua}{david.rios@icmat.es}{1}


\affiliation{1}{Institute of Mathematical Sciences}
  {ICMAT-CSIC}

\begin{abstract}
Real time large scale streaming data pose major challenges to forecasting, in particular defying the presence of human experts to perform the corresponding analysis. We present here a class of models and methods used to develop an automated, scalable and versatile system for large scale forecasting oriented towards safety and security monitoring. Our system provides short and long term forecasts and uses them to detect safety and security issues in relation with multiple internet connected devices well in advance they might take place.

\keywords Forecasting, Real time predictive analytics, Network monitoring, Dynamic models, Bayesian methods

\end{abstract}

%
%

\section{Introduction}
\label{sec:introduction}
 
As outlined in \cite{mortenson2015operational}, leveraging big data and real time analytics constitute two main research venues for OR/MS in the analytics age. Information and communication technologies (ICT) have experienced an exponential growth in the last few decades and most human activities, businesses and devices strongly depend on ICT \cite{ict}. With the advent of the Internet of Things (IoT), this interrelation will become even more evident and change dramatically the way in which different components of business and service systems interact.

In parallel, risks concerning the security of ICT systems are also growing,
 as pointed out  e.g.\ in \cite{tenkey}. Such cyber risks can be of natural origin or man-made, the latter potentially originating from human failure or intentional interests, as in cyber crime. This actually constitutes currently a major global threat, as reflected in the Global Risks Map 2018, \cite{wef2016}. As an example, \cite{cyberloss} catalogues around 70 new potential cyber threats per minute, and estimates the annual cost to the global economy from cyber crime to be above 400 billion USD. This impact highlights the need for developing solid cybersecurity risk management frameworks and the central role that these will play in business security in the near future. In addition, the increasing significance and availability of data in every business-related activity emphasizes data-driven approaches to improve cybersecurity decision making. As an example, \cite{cybersecurity2014framework} provides a set of standards and guidelines supporting such frameworks, including several key cybersecurity activities. Two of them refer to continuous safety monitoring of Internet Connected Devices (ICDs) and anomaly detection. This has led to the development of tools that periodically collect high frequency information from the ICDs of an organisation to support their monitoring. Given the increasing relevance of ICT, we may need to face organisations with several hundred thousands of ICDs from which we obtain tens of variables every few minutes. This poses tremendous challenges in processing such enormous amounts of data and making the relevant forecasts and decisions in real time to mitigate, sufficiently in advance, potential or actual safety and security issues in a network. In particular, it is virtually impossible to analyse each individual device time series through people, creating the need for an automated framework and system. Moreover, within this context, many classical time series analysis models become useless, either because they cannot tackle a huge amount of high frequency data, or because they cannot be used in an automated fashion, due to the versatility of the series that need to be faced.

We introduce in this paper a framework for time series monitoring and anomaly detection which serves as basis for a system for large scale safety and security network monitoring. Functionally, we require the system to be:
\begin{itemize}
\item {\em Automatic}. Given the huge amount of series to be monitored, intervention of humans in the  process should be kept to a minimum.
\item {\em Scalable}. The approach should be scalable both in time and memory space. It should be fast 
to be able to cope with many very high frequency time series. In addition, due to the huge amount of time series to be monitored, it should be able to summarize each series with just a few parameters, avoiding storage of the whole series.
\item {\em Versatile}. Time series may be of different nature and have different characteristics such as linear growth, seasonality or outbursts. The approach should be able to deal with all these particularities in an automated fashion. In addition, the time series features may change over time and, thus, the framework should be able to adapt fastly.
\end{itemize}

Earlier work in network monitoring does not fully cover the above requirements. In his classic paper   \cite{brutlag2000aberrant}, Brutlag proposed a simple approach based on Holt-Winters forecasting; however, model parameters need to be set and tuned for each model to work well, complicating automation. In addition, his technique is not capable of tracking several seasonal periods, reducing its versatility. In \cite{prophet}, the authors recently proposed an analyst-in-the-loop algorithm which makes use of human and automated tasks, precluding full automation. Moreover, they essentially frame the monitoring problem as a curve-fitting exercise using Generalized Additive Models, not fully taking into account the temporal dependence structure in the data. This way, the dynamic nature of the algorithm and its adaptability to sudden changes are essentially lost.  In \cite{vallis2014novel} a tool is presented for anomaly detection based on a seasonal trend loess decomposition together with the Generalized Extreme Student Deviation (GESD) test to detect anomalies. This algorithm handles anomaly detection issues, but does not accomplish other important monitoring tasks in relation with forecasting potentially dangerous future events. The main shortcoming is its rigidity in adapting to situations in which the signal features change considerably. Classification and regression trees (CART), \cite{breiman1984classification}, are also popular methods for anomaly detection. They may be used to provide point and interval forecasts as well as detect anomalies through GESD or Grubbs' tests; their main disadvantage is that a growing number of features can impact computational performance quickly, jeopardising scalability. ARIMA models have also been widely used in this area \cite{bianco2001outlier}. These models assume that the signal under study is stationary, possibly after differencing, which is not always the case in many network monitoring time series. In addition, numerous parameters such as the number of differences should be selected in advance, complicating automation. Finally, long short-term memory (LSTM) neural networks, if properly built, may perform successfully in anomaly detection tasks \cite{malhotra2015long}. However, their main drawback stems from the complex  work required to adjust them to reach a proper performance level, rendering automation virtually infeasible.

Our approach represents a step towards the achievement of a completely automated, scalable and versatile system for time series monitoring and anomaly detection in the specific area of network safety and security. In this domain, aberrant behaviour identification is often based on heuristic methods developed by analysts, and it usually lacks predictive capabilities. The framework we propose aims at providing such predictive power to the main monitoring system in a fully automated way.  For that purpose, it uses a Bayesian approach differentiating between continuous and discrete series. For continuous ones, we use a modified version of dynamic linear models (DLM) \cite{harrison1999bayesian}, that incorporates into the monitoring system the eventual existence of regular outbursts originated by physical processes such as backups or compressions, specially relevant in our application domain. A main advantage of DLMs is that they can be constructed using different blocks capturing each of the specific features of the time series. We provide an effective way to automatically identify the involved models. In addition, these models summarize the relevant aspects of the series in a few parameters, facilitating space scalability. Moreover, computation of the predictive distributions is relatively fast, thus making the approach time scalable after appropriate tunning. For discrete series, we use discrete time Markov chains \cite{insua2012bayesian}, which fulfill also our requirements above.

The structure of the paper is as follows. A description of the models and their goals are given in Section \ref{sec:modelDefinition}, together with how models are identified and how forecasts are made. This helps us in covering the versatility and automaticity requirements. We then discuss implementation details in Section \ref{sec:implementation},
covering scalability requirements. We end up with some concluding remarks.

\section{Problem formulation and model description}
 \label{sec:modelDefinition}

Consider, for the moment, the case of monitoring a single time series which, in our case, will refer to some performance measure of an ICD. Associated with the series, there are two reference values, designated warning ($W$) and critical ($C$), linked with observation levels such that, if exceeded, should lead to issuing warning and critical signals, respectively. For example, for a storage disk, we could be monitoring its usage and set $C=0.95$, meaning that when we reach a saturation of 95\%, disk performance might degrade and even collapse, inducing a loss. Such thresholds will depend on the device and its overall relevance. Observe that we have established a two-level alarm system in order to try to get richer information about potential failures of the monitored device. In the above example, $W = 0.9$ would allow us to raise awareness before it is too late.

As pointed out, several key cybersecurity activities are related with safety monitoring and anomaly detection within time series, which, in turn, we shall base on forecasting by:
\begin{itemize}
\item[i.] Making short term forecasts. These allow us to:

\begin{itemize}
\item Identify anomalous behaviour of the series when observed values do not lie within predictive intervals. This is related with security and could be useful in pointing critical issues in advance or detecting intruders in a system.

\item Identify unsafe behaviour when the predictive intervals cover the $W$ and/or $C$ thresholds.
\end{itemize}

\item[ii.] Making long term forecasts. These allow us to foresee critical issues sufficiently in advance when the $W$ and/or $C$ thresholds appear in the long-term predictive intervals: critical values that lie within them point to potentially problematic behaviour in the future.
\end{itemize}
To accomplish such tasks, we need procedures to make point and interval predictions of the involved time series. The width of the intervals should be adaptable to control for false positives, as well as to take into account the value of the corresponding asset. Whenever the predictions reach either level $W$ or level $C$, an alarm should be issued, being stronger in the latter case. In addition, we should emphasize alarms whenever several of them occur at consecutive time periods.

Depending on the specific nature of the series considered, a suitable model for each case becomes necessary. We first describe the models used for continuous time series network monitoring and, then, those for discrete time series.


\subsection{Continuous time series}

Typical continuous time series examples in our domain include device load and disk storage. For a given organisation, let $Z_n$ be the $n$-{th} observation of a relevant continuous variable monitored. If $h$ is the monitoring period, as configured by the user, $Z_n$ represents the observation at time $n \cdot h$. $D_n$ represents the values observed before such time and is recursively defined through $D_n = D_{n-1} \cup \lbrace z_n \rbrace$.

\subsubsection{Model Definition} \label{model_def_cont}
Based on intensive exploratory analysis of numerous time series in our domain, we have identified the following relevant types in high frequency traffic monitoring:
\begin{itemize}
\item Series with a level or linear trend, possibly varying in time.
\item Series with a level or linear trend together with one (or more) seasonal terms,
 typically describing daily and/or weekly variations.
\item Series with a level or linear trend, together with several outburst terms, typically associated with compression or backup processes.
\item Series with a level or linear trend together with one (or more) seasonal blocks as well as several outburst terms.
\end{itemize}

\begin{figure}
  \centering
    \includegraphics[width=0.7\textwidth]{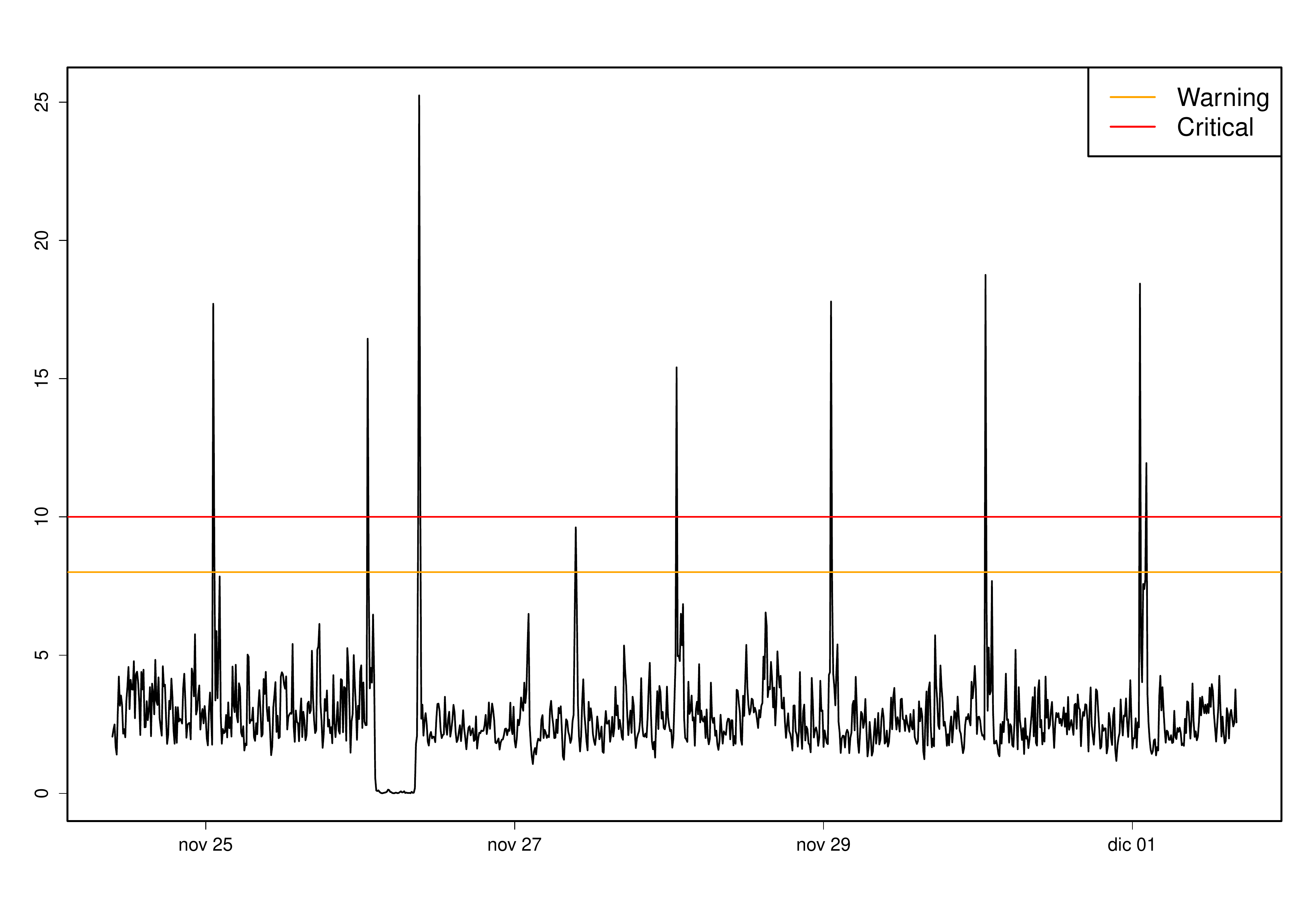}
      \caption{ Continuous time series with linear and outburst terms. \label{contSeries}}
\end{figure}
As an example, Figure \ref{contSeries} shows the average load of a device that goes through a backup process every night producing instantaneous load outbursts. This series would lie within the third type above. 	

As a consequence of the relevant types of series identified, the general expression that we shall adopt for our models will be
\[ Z_n = Y_n + S_n + B_n + \epsilon_n ,\]
where $Y_n$ designates a linear trend block; $S_n$ designates the seasonal block(s); $B_n$ designates the outburst block(s); and, finally, $\epsilon_n$ designates a noise term. Note that not all three blocks will need to be included in specific cases as explained here. We describe a procedure to automatically identify such blocks in Section \ref{modelid}. We outline first their definition.

The trend and seasonal terms are specifications of DLMs \cite{harrison1999bayesian}. We consider for them the general, normal DLM with univariate observation $X_{n}$, where $X_{n}$ corresponds either to the linear trend or the seasonal term. They are characterized by the quadruple $\{F_{n},G_{n},V_{n},W_{n}\}$: for each $n$, $F_{n}$ is a known vector of dimension $m\times1$, $G_{n}$ is a known $m\times m$ matrix, $V_{n}$ is a known variance, and $W_{n}$ is a known $m\times m$ variance matrix. The model is then succinctly written as
\begin{eqnarray*}
\begin{array}{L@{\quad}c}
Observation: & X_{n}|\theta_{n}    \sim N (F_{n}^{\prime}\theta_{n}, V_{n}), \nonumber \\
State: & \theta_{n}|\theta_{n-1}    \sim N (G_{n}\theta_{n-1}, W_{n}),  \nonumber \\
Prior: & \theta_{0}|D_{0}    \sim N (m_{0}, C_{0}),
\end{array}
\\
\end{eqnarray*}
where $\theta_n$ represents the state variable at time $n$. We specify now the general model for the required blocks, which we may combine using the superposition principle \cite[p.~186--188]{harrison1999bayesian}. The trend model is a specification of the DLM with constant $F_n$ and $G_n$ through
\begin{eqnarray*}
F = \left[\begin{array}{cc}
1&0
\end{array} \right],
\hspace{1cm}
G = \left[ \begin{array}{cc}
1&1\\
0&1\\
\end{array} \right].
\end{eqnarray*}
In turn, the basic seasonal model with period $s$ is a specification of the DLM with constant $F_n$ and $G_n$
\begin{eqnarray*}
F = \underbrace{\left[\begin{array}{cccc}
1&0&\cdots&0
\end{array} \right]}_{s-1},
\hspace{1cm}
G = \underbrace{\left[ \begin{array}{ccccc}
-1&-1&-1&\cdots &-1\\
1&0&0& \cdots &0\\
0&1&0& \cdots &0\\
\vdots& & && \vdots\\
0&0&\cdots&1&0\\
\end{array} \right]}_{s-1}.
\end{eqnarray*}
Typical periods that we incorporate are $s=288$ for 5 minute data and $s=144$ for 10 minute data. For larger periods, we might use a Fourier decomposition and work with a few of the Fourier components \cite[p.~102--109]{rbook}. For both models we use $V_n = 1$; however, this parameter could be assessed using e.g.\ maximum likelihood estimation.  The variance matrix $W_n$, is defined by a discount principle with discount 0.95 \cite[p.~193--200]{harrison1999bayesian}.


In case some outburst processes are detected, then at the corresponding times, the DLM model is \textit{turned off} and the analysis of these points is made separately. To this end we use a normal model for each particular outburst process
\begin{equation*}
B_{n_p} \sim \mathcal{N}(\mu , \sigma^2 ),
\end{equation*}
with mean $\mu$ and variance $\sigma^2$. Here, $n_p$  would represent the index of the $p$-th outburst of a given type.


\subsubsection{Model Identification} \label{modelid}

By default, we use as baseline a linear trend block on top of which we add seasonal and outburst blocks when such effects seem relevant. These are identified as follows:
\begin{itemize}
\item \textbf{Seasonal component:} We store the lags $t_i$ at which sign changes in the sample autocorrelation function of the series take place.  We then calculate the difference between the closest non-consecutive lags, $(t_{i+2} - t_i)$ for $i = \{ 1, \cdots , j-2 \}$, assuming that there are $j$ changes, and compute the mean ($m$) and variance ($s^2$) of differences. If $s / m < r$, where $r$ is an adjustable threshold, we include a seasonal component with nearest integer to $m$ as the estimated seasonality. Figure \ref{ACF} represents the autocorrelation of a series of period 144 in a specific case.

\begin{figure}[h]
  \centering
    \includegraphics[width=0.7\textwidth]{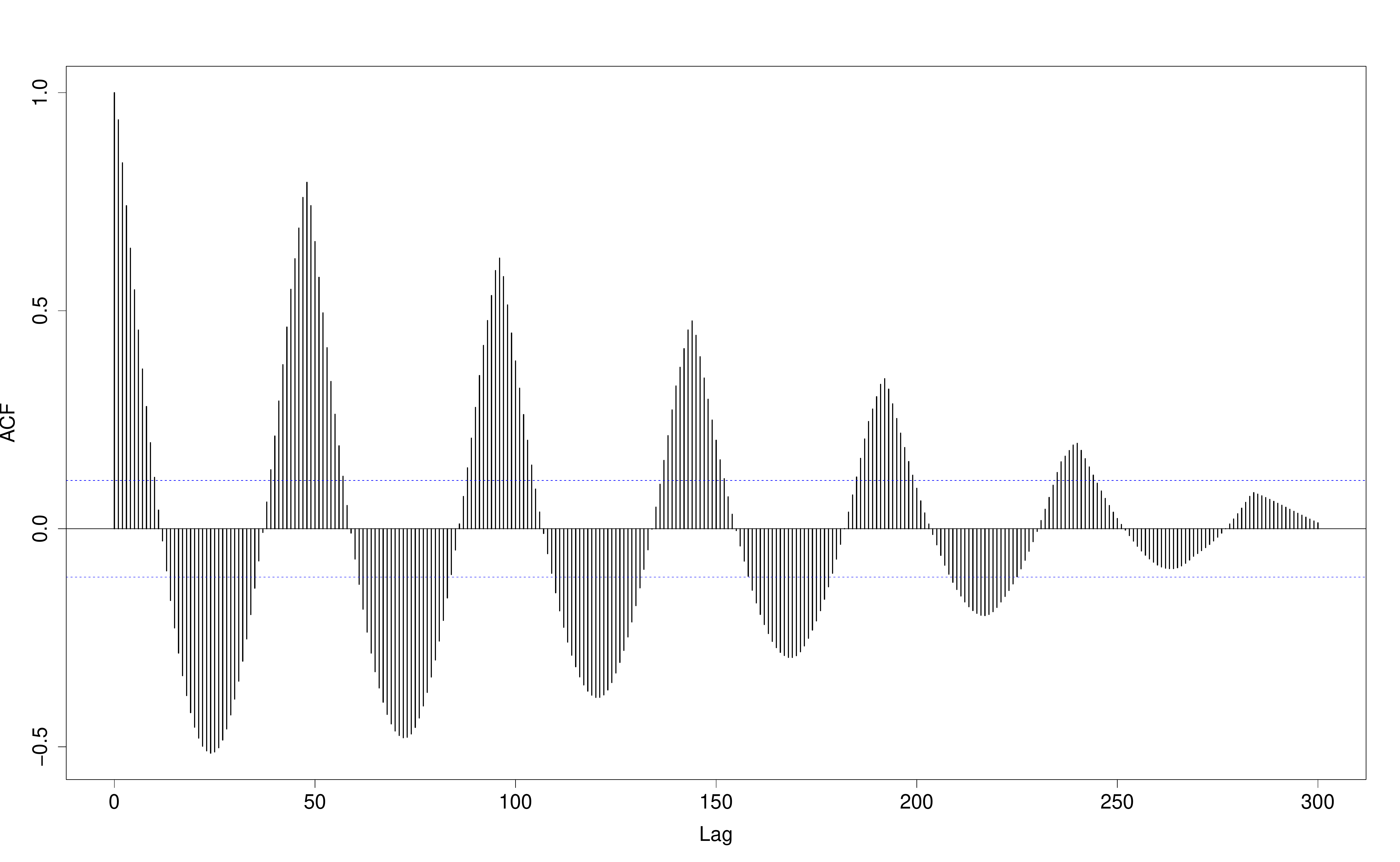}
      \caption{Sample autocorrelation function of a series with seasonal component.\label{ACF}}
\end{figure}

\item \textbf{Outbursts:} We first search through the dataset looking for times in which the data lie outside $\gamma$ standard deviations from the estimated mean of the series, with $\gamma$ an adjustable threshold. 
For these times we record how many instances $b_j$ of the same type of outburst occur, e.g\ for analyzing daily-regular outbursts we take a certain number of days and count in how many of them does a peak appear at the same hour. After a sufficiently long time, we declare that this is an identified regular outburst process if
\begin{equation}
\frac{b_j}{b} > q, \label{peakCondition}
\end{equation}
where $b$ is the number of periodical time intervals in the data used for model identification purposes, e.g.\ the total number of days in our previous example. $q$ is a \textit{repetition threshold} that can be tuned to make peak selection stricter. Condition \eqref{peakCondition} suggests that peaks need to appear, at least, $q \times b$ times throughout the data timespan to be considered as a part of a regular outburst process.
\end{itemize}
For this identification process, we use a large enough amount of data, able to capture the main features of the time series under study. In our application domain, where daily and weekly effects are the most relevant, 5 weeks is usually enough.
\subsubsection{Priors}
It could be the case that we have available prior information for specific monitored series. This will typically entail improved performance in terms of faster fitting. However, to automate the approach we need to be generic and we have adopted specific priors as follows:\\

{\em Linear term}.  The adopted prior is $\mathcal{N}(m_0, C_0)$, where
\begin{equation*}
\begin{array}{cc}
\begin{aligned}
    m_0 &= \begin{bmatrix}
          0 & 0 \\
         \end{bmatrix}
  \end{aligned}, & 
   C_0 =
\begin{bmatrix}
10^7 & 0 \\
 0  & 10^7
\end{bmatrix}.
\end{array}
\end{equation*}

{\em Seasonal term}.  We use again an $\mathcal{N}(m_0, C_0)$ prior. $s$ being the period, we adopt as $m_0$ a vector of 0's of dimension $s-1$. The covariance matrix, of dimension $(s-1) \times (s-1)$, takes the form:
\begin{equation*}
C_0 = \begin{bmatrix}
A & B & \cdots & B\\
B & \ddots &  & \vdots\\
\vdots &  & \ddots & B\\
B & \cdots & B  & A
\end{bmatrix},
\end{equation*}
where $A$ is a large positive value and $B$ is a negative value defined so that the sum of every row and column is zero. Thus, we set $B = - \frac{A}{s-2}$.\\

{\em Outburst term}.  We use a noninformative prior
\begin{equation*}
\Pi (\mu, \sigma^2) \propto \frac{1}{\sigma^2}.
\end{equation*}

\subsubsection{Model Forecasting} \label{ModFor}
We describe how we make forecasts with the above models. In \cite[p.~53--55]{rbook} it is provided the closed form of the one-step ahead predictive distribution $\mathcal{N}(f_n, Q_n)$ of $X_{n}|D_{n-1}$, affecting both the trend and seasonal terms, and their superposition thereof.
As a consequence:
\begin{itemize}[leftmargin=*]
\item The pointwise forecast at the next period $n$ is $E[X_n|D_{n-1}] = f_n$.
\item If $u_n$ and $l_n$ represent, respectively, the upper and lower bounds of the predictive interval, they are defined through
\begin{eqnarray} \label{interval}
u_n &=& f_n + z_{1-\alpha/2}\:Q_n^{1/2}, \nonumber \\
l_n &=& f_n - z_{1-\alpha/2}\:Q_n^{1/2},
\end{eqnarray}
where $z_{1-\alpha/2}$ is the $1-\alpha/2$ quantile of the standard normal distribution, being $\alpha$ the desired probability level of the predictive interval, which may be chosen by design, e.g.\ depending on the asset value of the corresponding device. By default, we use $\alpha = 0.95$.
\end{itemize}

$k$-step ahead forecasts are also relevant in our context, both for short and long term. If we use $a_n(k) = E[\theta_{n+k}| D_n]$, $f_n(k) = E(Y_{n+k}|D_n)$, and $Q_n(k) = Var(Y_{n+k}|D_n)$, \cite[p.~70--71]{rbook} provide closed forms for such means and variances.
%
As a consequence:
\begin{itemize}
\item The $k$-step ahead point forecast for the trend term at time $n$ is
\begin{eqnarray}
f_n(k) = a_n(0)+a_n(1)k. \label{kforelin}
\end{eqnarray}
\item The $k$-step point forecast for the seasonal term at time $n$ is
\begin{eqnarray}
f_n(k) = a_n(0)_{k \text{ mod } s}. \label{kforeseas}
\end{eqnarray}
\end{itemize}
Prediction intervals take a similar form as in \eqref{interval}.

For the outbursts, for simplicity, let us index the peaks of a particular type with $p$, the number of observed outbursts of such type until the corresponding time.  Then

\begin{equation*}
B_{p+1}|D_{p} \sim \mu_p + t_{p-1}\sqrt{\left( 1 + \frac{1}{p} \right) \sigma_p^2},
\end{equation*}
where $\mu_p$ is the sample mean after observing $p$ outbursts of the type being studied, and $\sigma_p^2$ is the sample variance. We may update  recursively the parameters through
\begin{equation*}
\begin{aligned}
\mu_{p+1} &= \frac{p \mu_p + X_{p+1}}{p+1}, \\
\sigma_{p+1}^2 &= \frac{\sigma^2_p (p-1) + k \mu_p^2 + X_p^2 - (p+1)\mu_{p+1}^2}{p},
\end{aligned}
\end{equation*}
where $X_{p}$ is the value of the series at the time corresponding to the $p$-th outburst of the given type. Then,
\begin{itemize}
\item The {\em point} forecast is $E[B_{p+1}|D_p] = \mu_p$.
\item The {\em interval} forecast is
\begin{equation}\label{outInt}
\begin{aligned}
u_{p+1} &= \mu_p + t_{1-\frac{\alpha}{2}, p+1} \sqrt{\left( 1 + \frac{1}{p} \right) \sigma_p^2}, \\
l_{p+1} &= \mu_p - t_{1-\frac{\alpha}{2}, p+1} \sqrt{\left( 1 + \frac{1}{p} \right) \sigma_p^2},
\end{aligned}
\end{equation}
where $t_{1 - (\alpha /2 ), p+1}$ is the $1-\alpha/2 $ quantile of the $t$-distribution with $p-1$ degrees of freedom, with $\alpha$ as above.
\end{itemize}

Finally, for general forecasts, we first distinguish between regular points and outbursts. For regular points, the prediction is based on the superposition of the involved DLM components. However, if the prediction refers to an outburst, the DLM model is switched off and the outburst forecast is used.

\subsection{Discrete time series}
\label{modelDiscrete}
A typical discrete time series examples in our domain include the number of petitions to a server. We briefly sketch the approach with this type of series. For further details see \cite{insua2012bayesian}. We use finite, time homogeneous Markov chains $\lbrace X_n \rbrace$, with states $\{ 1,\ldots,K\}$, where $K = C+c$, is the stated critical level plus a small integer $c$. We write the transition matrix as ${\bf P} = (p_{ij})$ where $p_{ij} =P(X_n=j|X_{n-1}=i)$, for $i,j \in  \{1,\ldots,K\}$. Should it exist, the stationary distribution ${\bf \pi}$ is the unique solution of ${\bf \pi} = {\bf \pi}{\bf P}$, $\pi _i \geq 0,\sum \pi_i=1$.

We assimilate the monitored data $D_n$ to observing $n$ successive transitions of the Markov chain, say $X_1=x_1,\ldots,X_{n}=x_{n}$, given the known initial state $X_0=x_0$. 
A natural prior for ${\bf P}$ is defined by letting $\textbf{p}_i = (p_{i1},\ldots,p_{iK})$ have a Dirichlet distribution
$ \textbf{p}_i \sim \text{Dir}\left( \pmb{\alpha}_i \right)$ 
where $\pmb{\alpha}_i = (\alpha_{i1},\ldots, \alpha_{iK})$
for $i=1,\ldots,K$.  
Then, 
the posterior is ${\bf p}_i|D_n \sim \text{Dir}\left( \pmb{\alpha}_i^{\prime} \right)$ where $\alpha_{ij}^{\prime} = \alpha_{ij}+n_{ij}$ for $i,j = 1, \ldots, K$; being $n_{ij} \ge 0$ the number of observed transitions from state $i$ to state $j$. Specifically, the prior adopted is given by the coefficients of the corresponding Dirichlet distributions presented in the following matrix of dimension $K \times K$
\begin{equation*}
  \begin{bmatrix}
\alpha & \beta & \gamma & \gamma & \cdots & \gamma \\
\beta & \alpha & \beta &  & \cdots & \gamma\\
\gamma & \beta &  \ddots     &         &  & \vdots \\
\vdots & \vdots &      &		   &   &   \beta  \\
\gamma &   \gamma     &    \cdots   &    & \beta     & \alpha \\
  \end{bmatrix}.
\end{equation*}
In our case, due to the high frequency of the time series involved, the most-likely behaviour for a certain state is to remain in its position, followed by one-state transitions. The rest of transitions are less likely. We model this behaviour by setting $\alpha > \beta > \gamma$. In particular, our default values are $\alpha = 10$, $\beta = 8$ and $\gamma = 2$. As before, the proposed prior is general, flexible and useful to deal with the cases we have found in network monitoring, facilitating a generic automated approach.

We predict the next value of the Markov chain, at
time $n+1$, using our estimates of the transition probabilities

\begin{eqnarray*} P(X_{n+1}=j|D_n) =   \int
p_{x_n j} f({\bf P}|D_n) \, d{\bf P}  =  \frac{\alpha_{x_n
j}+n_{x_n j}}{\alpha_{x_n \bullet}+n_{x_n \bullet}} \equiv \widehat{p}_{nj}, \end{eqnarray*}
where $\alpha_{i \bullet} = \sum_{j=1}^K \alpha_{ij}$ and $n_{i \bullet} = \sum_{j=1}^K n_{ij}$. The pointwise prediction will then be $\sum_{j = 1}^K j \cdot \widehat{p}_{nj} $. Once the prediction of the next value is performed, we can also calculate the prediction interval around such  estimation. 
\begin{algorithm}[h] 
 \KwData{Last visited state: $i$. Transition probabilities: $\widehat{p}_{i,j}$ for $j = 1,2,\dots, K$.}
 \textbf{Initialization:} \
 $~\text{sum} = p_{i,i}$, $~u_{n+1} = i$,$~l_{n+1} = i$ \;
 \While{$\text{sum} < \alpha$}{
 \eIf{$~u_{n+1} = K$}{
  continue, interval stops growing upwards\;
 }{
  $~u_{n+1} = u_{n+1} + 1$\;
  $~\text{sum} = \text{sum} + \widehat{p}_{i,u_{n+1}}$\;
 }
 \eIf{$~l_{n+1} = 1$}{
  continue, interval stops growing downwards\;
  }{
  $~l_{n+1} = l_{n+1} - 1$\;
  $~\text{sum} = \text{sum} + \widehat{p}_{i,l_{n+1}}$\;
  }
 }
 \KwResult{$u_{n+1}$, $l_{n+1}$}
 \vspace{2.5mm}
 \caption{Predictive interval calculation for discrete time series.} \label{algg}
\end{algorithm}
If $i$ is the last visited state and $\alpha$ is the probability of the one step ahead predictive interval, we get its upper and lower bounds $u_{n+1}$ and $l_{n+1}$, using Algorithm \ref{algg}. Notice that discrete time series have both \textit{maximum} and \textit{minimum} states, here $K$ and $1$ respectively. 

%
%
%

\noindent $k>1$ steps ahead predictions are more complex. For small $k$, we can use
\[ P(X_{n+k}=j|D_n) = \int \left({\bf P}^k\right)_{x_n j} f({\bf
P}|D_n) \, d{\bf P},\] which gives a sum of Dirichlet expectation terms. However, as $k$ increases, the evaluation of this expression becomes computationally infeasible in our domain.
%
Thus, we perform approximately by calculating the $k^{th}$ power of the matrix of expected values of the probabilities, $\widehat{P}^k$, and using
\begin{eqnarray*}
P(X_{n+k} = j| D_n) \approx \left(\widehat{P}^k\right)_{nj}.
\end{eqnarray*}

Our interest also lies in the stationary distribution of the chain. For high dimensional chains as the ones we need to deal with, 
we use the approximation
\begin{equation}
\begin{gathered}
\hat{\pi} = \hat{\pi}\hat{\mathbf{P}}, \\
\sum_i \hat{\pi}_i = 1, \\
\hat{\pi}_i \geqslant 0.
\end{gathered} \label{forecasting_discrete}
\end{equation}
 Once this equation is solved, we can use $\hat{\pi}_i \geqslant 0$, the approximate stationary distribution, to produce long-term forecasts.


\section{Implementation}
\label{sec:implementation}

This section describes the implementation and use of the approach presented in Section \ref{sec:modelDefinition}, which referred to a single monitored series focusing on versatility and automation. Here we describe aspects in relation with having to deal with several hundred thousands of monitored series sampled over periods ranging from 1 to 10 minutes, depending on the criticality of the corresponding device. We thus cover the speed and space scalability features mentioned above.

\begin{figure}[h]
\begin{center}
\begin{tikzpicture}[->, >=stealth', auto, semithick, node distance=3cm]
\tikzstyle{every state}=[fill=white,draw=black,thick,text=black,scale=1]
\node[state]    (t)   at(0,0)   {TS};
\node[state]    (c) at(2,0)   {Classifier};
\node[state]    (d) at(2,2.5)   {Discrete};
\node[state]    (cn) at(2,-2.5)   {Continuous};
\node[state]    (m) at(6,2.5)   {Markov};
\node[state]    (l) at(5,-1)   {Lin};
\node[state]    (s) at(5,-2.5)   {Seas};
\node[state]    (o) at(5,-4)   {Otbrst};
\node[state]    (dm) at(6.5,-2.5)   {DM};
\node[state]    (fit) at(7,0)   {Fit};
\node[state]    (f) at(9,0)   {Forecast};
\node[state]    (a) at(12,0)   {Alarms};
\path
(t) edge[bend left = 1]     node{}     (c)
(c) edge[bend left = 1]     node{}     (d)
(c) edge[bend left = 1]     node{}     (cn)
(d) edge[bend left = 1]     node{}     (m)
(m) edge[bend left = 1]     node{}     (fit)
(fit) edge[bend left = 1]     node{}     (f)
(f) edge[bend left = 1]     node{}     (a)
(cn) edge[bend left = 1]     node{}     (l)
(cn) edge[bend left = 1]     node{}     (s)
(cn) edge[bend left = 1]     node{}     (o)
(l) edge[bend left = 1]     node{}     (dm)
(s) edge[bend left = 1]     node{}     (dm)
(o) edge[bend left = 1]     node{}     (dm)
(dm) edge[bend left = 1]     node{}     (fit);
\end{tikzpicture}
\end{center}
\caption{General scheme.}\label{genSc}
\end{figure}
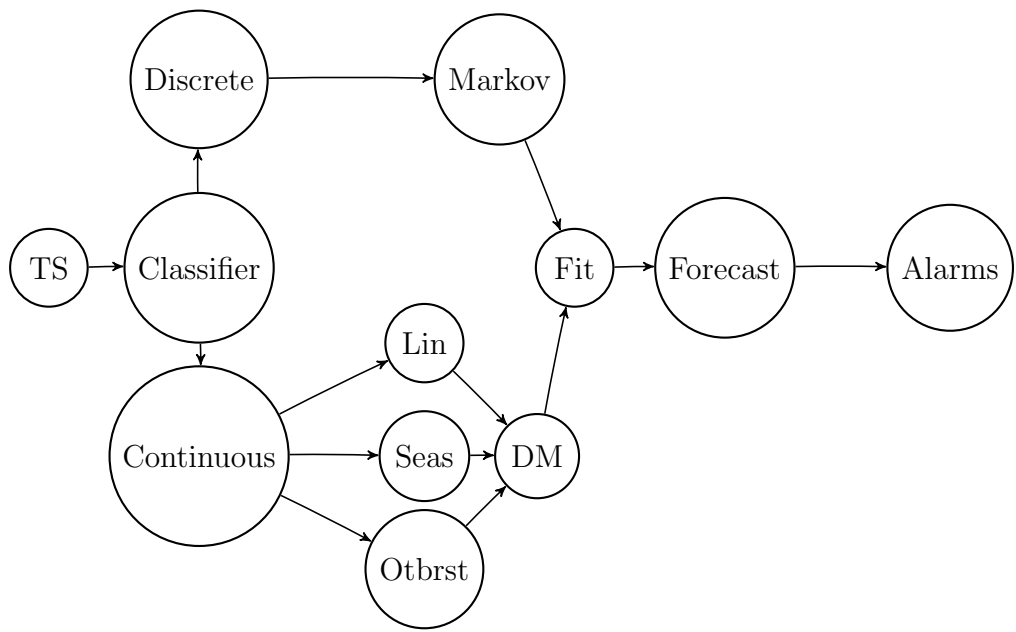

The general strategy followed for each time series is described in Figure \ref{genSc}. Once the series is incorporated into the monitoring system, a first distinction is made between whether it is continuous or discrete. In this last case, the system uses the Markov model in Section \ref{modelDiscrete}. Otherwise, the model identification process presented in Section \ref{modelid} is implemented. Once the blocks have been identified, the appropriate dynamic model in Section \ref{model_def_cont} is used. After fitting the corresponding models based on a sufficient amount of data, forecasts may be made and alarms raised in case an anomalous behaviour is detected or reaching critical values is predicted, either in the short or medium terms.

\begin{figure}[h]
\begin{center}
\begin{tikzpicture}[->, >=stealth', auto, semithick, node distance=3cm]
\tikzstyle{every state}=[fill=white,draw=black,thick,text=black,scale=1]
\node[state]    (D)   at(-3,0.7)   {Data};
\node[state]    (T) at(-3,-0.7)   {Time};
\node[state]    (C) at(2,1)   {Classifier};
\node[state]    (W) at(0,0)   {Main};
\node[state]    (M)   at(3,-3)   {Models};
\path
(D) edge[bend left = 1]     node{}     (W)
(T) edge[bend left = 1]     node{}     (W)
(C) edge[bend left]     node{\textbf{classifier}}     (W)
(W) edge[bend left]     node{\textbf{data,time}}     (C)
(W) edge[bend left = 1]     node{\textbf{data,time,classifier}}     (M)
(M) edge[bend left]     node{\textbf{gen}}     (W);
\end{tikzpicture}
\end{center}
\caption{Structure of the python package for time series monitoring and anomaly detection.}\label{pp}
\end{figure}
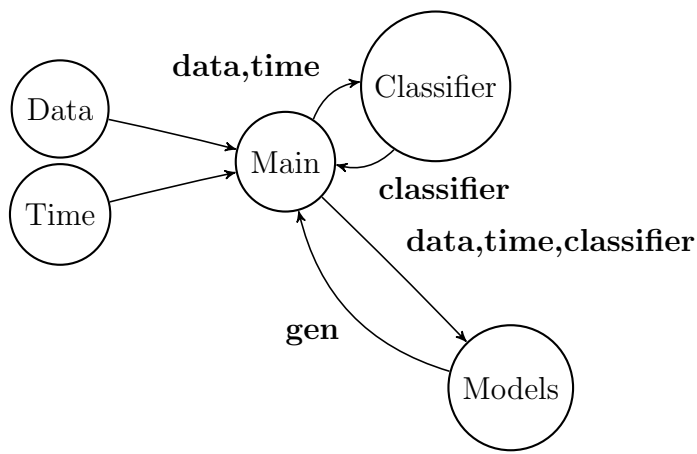

The implementation has been carried out through the creation of an object-oriented python package linked with a core monitoring system. Each time series monitored will be associated with a particular object that includes all methods required to accomplish predictive monitoring tasks. The package is structured as in Figure \ref{pp}.
The main code blocks are the classes \textit{gen} and \textit{classifier}. The main code receives the data and timestamps of the training period of each series and generates a \textit{classifier} object. Its attributes contain the characteristics needed to identify the class of models to be used in the analysis of that particular time series. Once this object is constructed, using it together with the data and timestamps, a  \textit{gen} object is created. This is the general class whose methods depend on the model relevant for the particular time series, which will ultimately be used to undertake the forecasting, anomaly and critical value detection. Among these, we highlight the  \textit{update} method, which incorporates new data to the current model and updates the parameters in a Bayesian manner, and the \textit{predict} method, which performs $k$-step ahead predictions, returning point and interval forecasts, providing the basis to construct the specific critical and anomaly detection functions that raise the alarms.

Short term forecasts are needed for safety and security purposes. Given $D_n$, we produce forecasts using the \textit{predict} method for future observations $x_{n+1}$, $x_{n+2}$, $x_{n+3}, \cdots$ up to a certain time $(n+k)\cdot h$, with $k$  defined by the user. $k=3$ is the default value. Forecasts are delivered as predictive intervals $[l_{n+i}, u_{n+i}]$ with probability level  configurable by the user. These could be used to detect aberrant behaviour within the time series: when the next observation $x_{n+1}$ does not lie in the predictive interval $[l_{n+1}, u_{n+1}]$, a warning of unexpected behaviour is displayed pointing to a potential security issue. Our implementation modulates the warning depending on the number of consecutive intervals in which this warning needs to be launched. So as to control for false positives, alarms could be issued just when the number of violations exceeds certain threshold within a moving window of  fixed number of time steps, as suggested in \cite{brutlag2000aberrant}. Short term predictions also serve for safety monitoring tasks. Specifically, when $W$ or $C$ lie within a predictive interval $[l_{n+j}, u_{n+j}]$ for one $j \in \lbrace 1, 2, ..., k\rbrace$, an alarm pointing to potentially high value of $x$ should be issued. The higher the number of intervals covering the particular level, the more intense would be the alarm.

\begin{figure}[htbp]
\subfloat[Continuous time series.]{\includegraphics[width=0.5\textwidth]{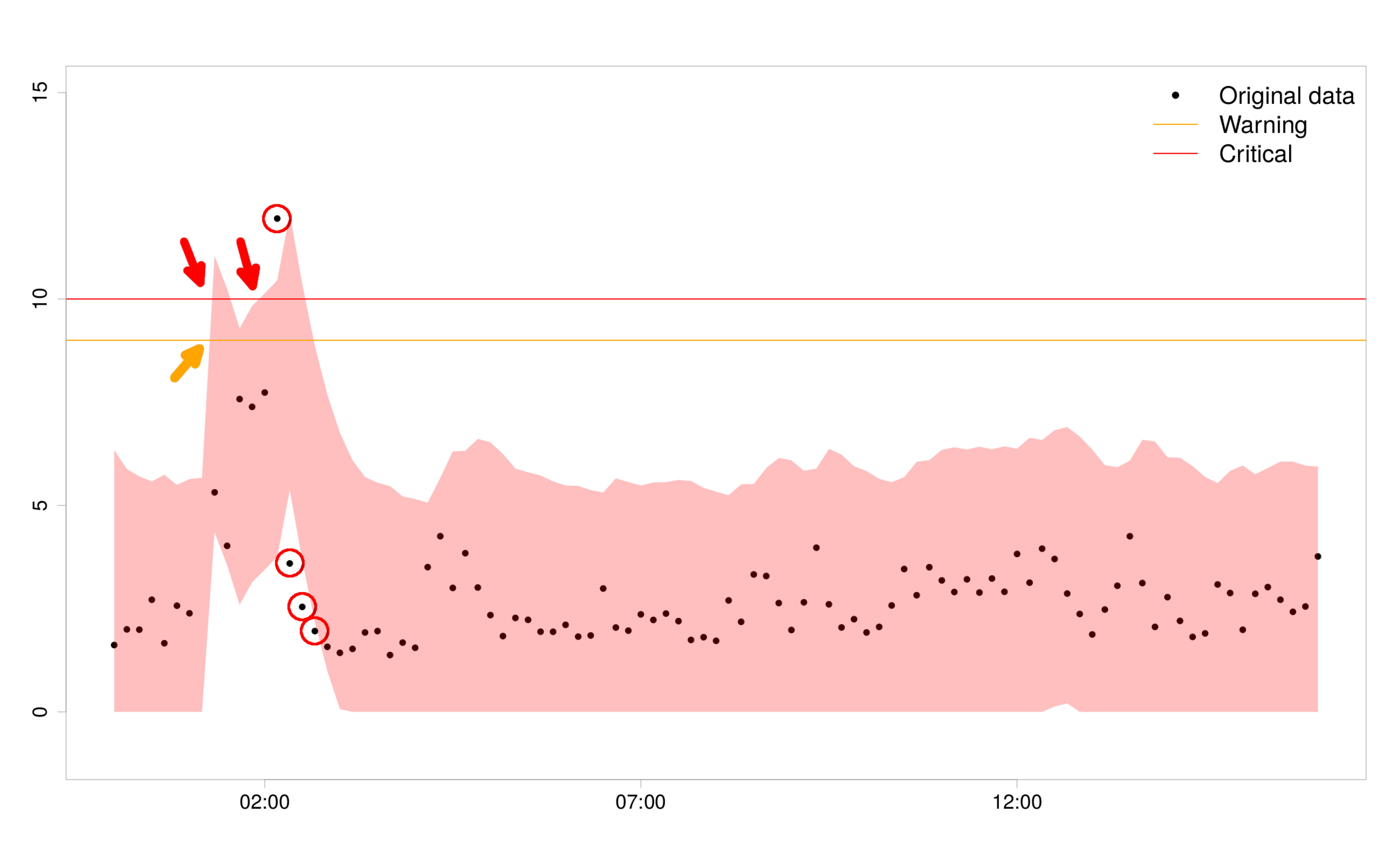}}
\subfloat[Discrete time series.]{\includegraphics[width=0.5\textwidth]{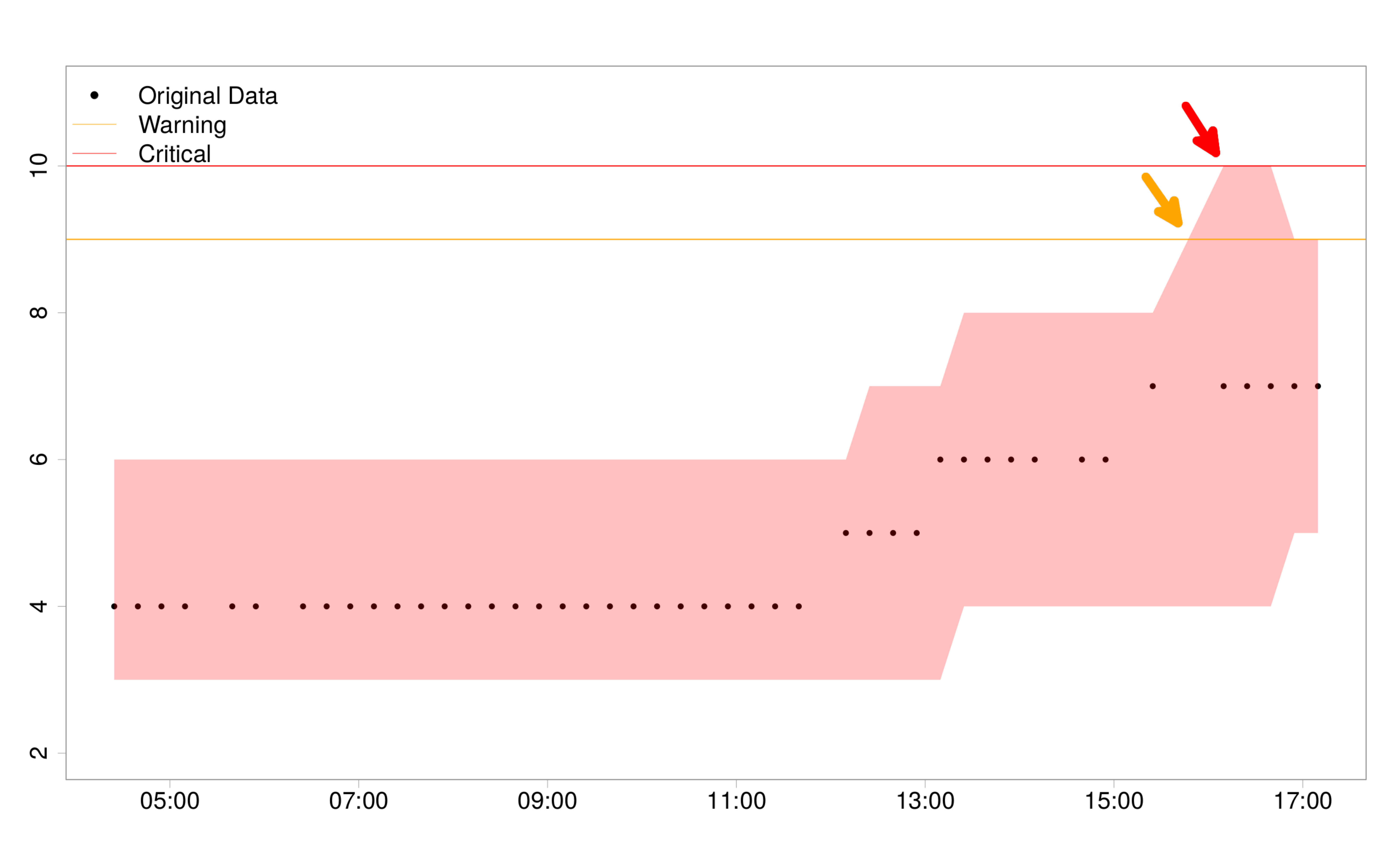}} \label{a}
\caption{Short term forecasting using 95\% one-step ahead predictive intervals.} \label{short_term}
\end{figure}

\noindent We illustrate both functionalities through a practical example in Figure \ref{short_term}(a).
In this case, an intense unexpected behaviour alarm would be issued around 2 AM as there are four measurements  outside the corresponding predictive intervals. In addition, an alarm concerning a potentially high value of the series would be issued at 1:40, because the predictive intervals exceed the warning and critical levels at 1:50 and 2:00, respectively. A similar example using discrete time series is in Figure \ref{short_term}(b).

Long term forecasts are used to accomplish safety monitoring tasks: we try to ascertain whether critical levels will be reached with sufficiently high probability in the long term. In principle, it could be done using predictive intervals as we do with short term predictions and illustrated in Section \ref{ModFor}. However, since the system must monitor hundreds of thousands of high frequency time series in real time and, consequently, must perform under a very narrow time window, we need to make a compromise: although the calculation of predictive intervals is doable as in Section \ref{ModFor}, long term we will just  focus on point forecasts. This allows for a drastic improvement in the computational costs of running the involved algorithms, as we can use explicit expressions to make the forecasts (and  obviate the costly computation
of predictive variances): we use as point forecasts $z_{n+j}$, the midpoint of intervals $[l_{n+j}, u_{n+j}]$, and try to find out the first $j_1$ such that $z_{n+j_1}>W$ and the first $j_2$ such that $z_{n+j_2}>C$. This allows us to identify, well in advance, time instants in which critical values might be reached, as we may provide explicit expressions of such inequalities. Indeed, in the linear case, from \eqref{kforelin}, we try to find the first $j_i, i =1,2 $ such that
\begin{eqnarray*}
\begin{array}{cc}
j_1 > \frac{W-a_n(0)}{a_n(1)}, &  j_2 > \frac{C-a_n(0)}{a_n(1)},
\end{array}
\end{eqnarray*}
assuming that $a_n(1) >0$. Similarly, with a seasonal block, from \eqref{kforeseas}, we try to find the first $j_i, i=1, 2$ such that
\begin{eqnarray*}
\begin{array}{cc}
a_n(0)_{j_1 \text{ mod } s} > W, &  a_n(0)_{j_2 \text{ mod } s} > C,
\end{array}
\end{eqnarray*}
if any. Finally, when the model contains linear and seasonal blocks, we compute $j_1 $ (and similarly $j_2$) through 

\begin{algorithm}[H]
 \While{$a_n(0) + a_n(1)k + a'_n(0)_{k \text{ mod } s} < W$}{
 $~k = k+1$;\\
 $~j_1 = k$;
 }
\end{algorithm}

\noindent where $a'_n(0)$ are the parameters of pointwise $k$-step ahead forecast for the seasonal trend, and $a_n(0)$, $a_n(1)$ are those of the linear trend.

\begin{figure}[h]
  \centering
    \includegraphics[width=0.7\textwidth]{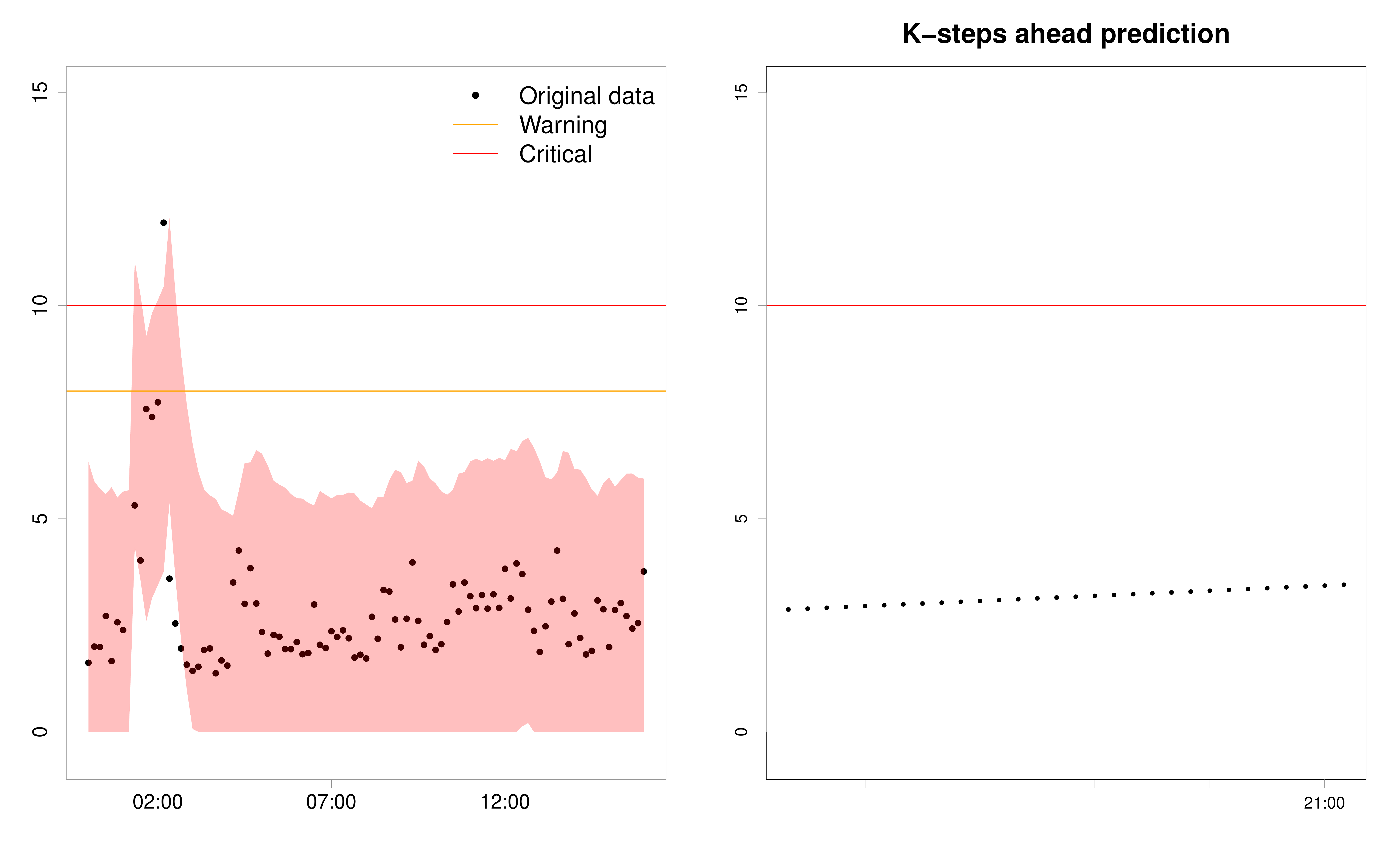}
      \caption{Long term forecast for continuous time series \label{long_term}}
\end{figure}

\noindent A practical example of this type of forecast is illustrated in Figure \ref{long_term}. There, the time series is not expected to cross neither warning nor critical levels in the next five hours. However, if the linear growth in the main trend does not change, both levels could be reached. In order to control for false positives we could define a \textit{relevance time window}, only raising an alarm when anomalous behaviour occurs inside it and neglecting them, otherwise.

\begin{figure}[h]
  \centering
    \includegraphics[width=0.7\textwidth]{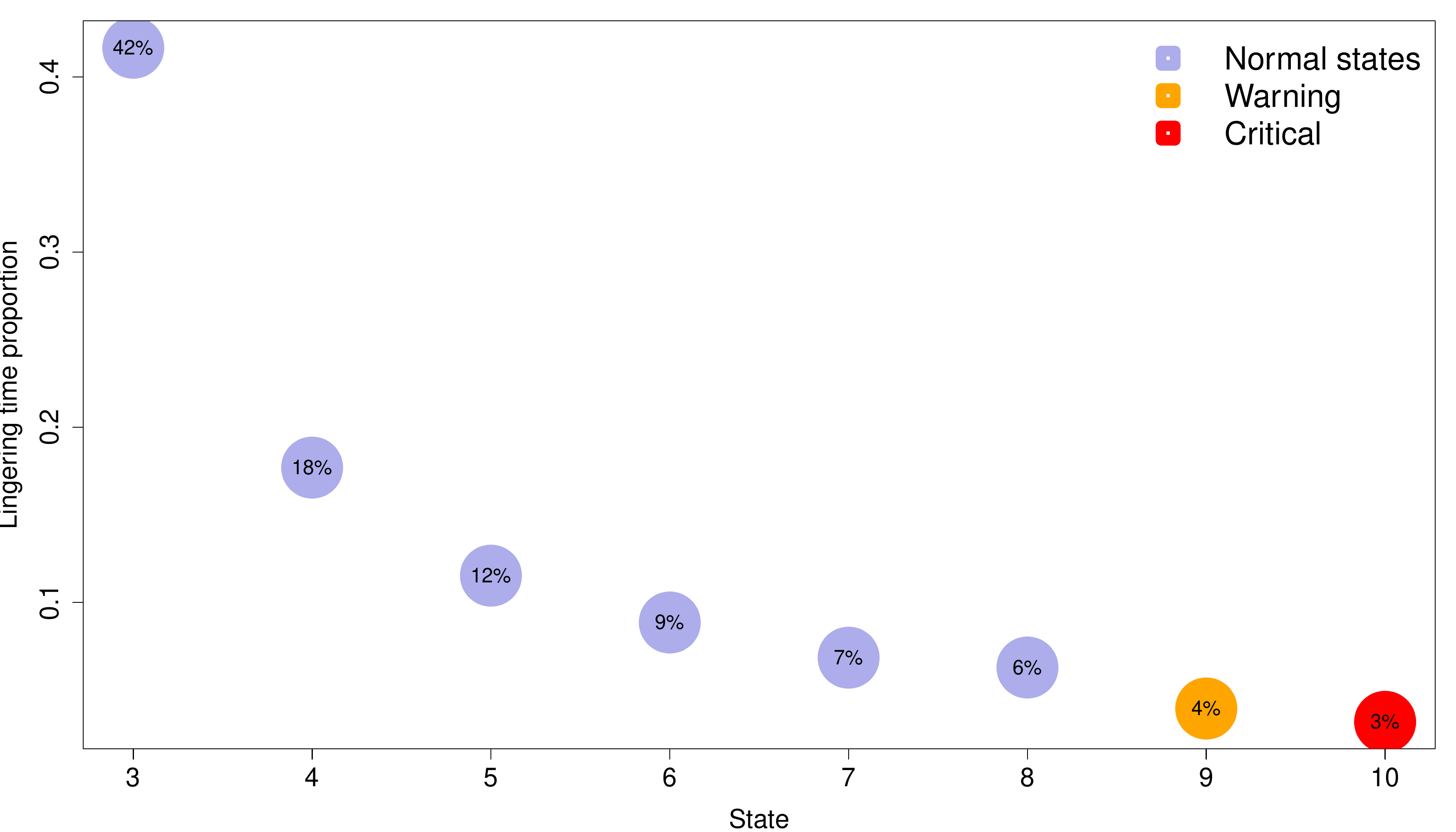}
      \caption{Long term forecast for discrete time series\label{long_term_discrete}}
\end{figure}

For long term forecasting with discrete time series, we use the stationary distribution in \eqref{forecasting_discrete}. If the sum of probabilities assigned to the states that lie above the warning (critical) level is higher than a certain prestablished threshold, an alarm would be raised. This is illustrated in Figure \ref{long_term_discrete}. In this case, if the warning (critical) threshold is at 0.07 or lower, an alarm should be raised. In any other case, the system would continue without issuing warning.

We assessed the time performance of our approach with a stress test. For different models, we have sequentially added 15000 new data points through the \textit{update} method, performing a short term prediction (30 minutes) and a long term prediction (5 hours) at each iteration. In particular, we tested four models: a model with a linear trend; one with a linear trend and a seasonal block with period 144; a model with a linear trend and an outburst block; and, finally, a Markov chain model with 50 states. In Table \ref{tab:times}, we show the mean, median, min and max times of the whole operation (update + short term prediction + long term prediction), for each of the models used. Time calculations have been performed in an Intel Core i7-3630UM, $2.40 GHz \times 8$ machine.
\begin{table}[ht]
\centering
\begin{spacing}{0.2}
\begin{tabular}{lcccc}
\toprule
 
                          & \textbf{Mean}      & \textbf{Median}     & \textbf{Min} & \textbf{Max}        \\ \midrule
Linear & $4.03\cdot 10^{-4}$ & $3.97\cdot 10^{-4}$ & $3.76\cdot 10^{-4}$ & $9.12\cdot 10^{-4}$ \\ \midrule
Linear + Seasonal  & $2.50\cdot 10^{-2}$ & $2.49\cdot 10^{-2}$ & $2.43\cdot 10^{-2}$ & $4.60\cdot 10^{-2}$ \\ \midrule
Linear + Outburst & $4.42\cdot 10^{-4}$ & $4.35\cdot 10^{-4}$ & $2.73\cdot 10^{-4}$ & $8.38\cdot 10^{-4}$ \\ \midrule
Markov Chain & $8.81\cdot 10^{-4}$ & $8.06\cdot 10^{-4}$ & $7.70\cdot 10^{-4}$ & $1.96\cdot 10^{-3}$ \\ 
\bottomrule
\end{tabular}
\end{spacing}
\vspace{0.4cm}
\caption{Mean, median, min and max times in seconds, of update, short term and long term prediction for different models.}\label{tab:times}
\end{table}

\noindent The algorithm is fast enough, and consequently, able to cope with the typical very high frequency data of most security domains. Note though that the model including the seasonal trend is remarkably slower than the rest of models. This is to be expected since it includes a seasonal term with high period (144), which involves performing several operations with high dimensional matrices. Should better performance be required, we could use a Fourier decomposition of the seasonal trend, and work with a few of the most relevant Fourier components \cite[p.~102--109]{rbook}, although this would require a method to automate identification of such components.

In terms of memory, the whole approach is constructed so that a fixed number of parameters are stored at any step of the calculations to further improve its performance. In the case of continuous time series, the linear part of the DLMs stores two 2-component vectors, the vector $F$ and the mean of the state distribution $m$; three $2\times 2$ matrices, the matrix $G$, the covariance matrix $C$ of the state distribution and the system covariance; and a parameter corresponding to the observation variance.  Similarly, the seasonal part stores two $(s-1)$-component vectors, three $(s-1)\times (s-1)$ matrices and one parameter. The outburst term holds only two parameters corresponding to the mean and variance of the outburst distribution, for each type of peak detected. Finally, the Markov model for discrete time series employs only a $k\times k$ matrix that is updated at each step, where $k$ corresponds to the number of states considered in the chain. This all means that in the worst case scenario, the model needs to store in memory $\mathcal{O}(s^2)$ parameters for the continuous case and $\mathcal{O}(k^2)$ for the discrete case, which is feasible in the cases being analyzed. These remarks, together with the time performance measures previously presented, show that the algorithm fulfills the time and space scalability request.




\section{Discussion}
 \label{sec:discussion}

The framework presented is very flexible when it comes to identify and model different behaviours that can be described in terms of basic components, as it happens in our application domain of network monitoring. This allows for a wide use through very different environments beyond our application domain, as we may use the same approach to monitor very distinct time series with varying intrinsic nature. Moreover, since the procedure just needs to store few parameters for each time series, we can say that it is scalable, both in terms of memory and runtime. The framework is amenable of parallel processing, making possible to monitor different batches of series in different cores, reinforcing this way the scalability request. In addition, thanks to our model identification procedure, our approach is able to work automatically with no human supervision. Another interesting advantage refers to updating series  with missing data, which relatively frequently happens in our application domain. When there are missing values in the dataset, then the state carries no information and, therefore, the filtering distribution at such time is just the one-step-ahead predictive distribution of the previous time. The framework has been implemented and operates as part of a system controlling a network of more than three hundred thousand devices, even taking into account that each of them provides several very high-frequency time series.

Improvement of the algorithm can still be done, specially if we take into account specifics of the cases monitored. To this extent, using the possible hierarchy between monitored devices, model identification could be optimized, for example allowing for a faster regular outburst detection. Moreover, should there exist correlated time series in the database, these correlations could be used similarly to our advantage.

It is also possible to carry out an automatic performance evaluation of the algorithm. This could be done by computing the autocorrelation function of the residuals within a certain time-window. If for instance, a strong correlation in the first few lags is encountered, an autoregressive term could be added to the model. In other cases, when model performance deteriorates too much, an alarm could be issued, demanding the intervention of an analyst.

It may be also interesting to adjust in real time the width of the probability intervals taking into account the particular potential economical losses of false positives and negatives of the system. From this point of view, we may also find interesting to re-build the structure of the algorithm to try to produce long-term forecasts including the probability intervals as well, moving beyond the point-wise predictions here proposed for practical implementation.


\section*{Acknowledgments}
RN acknowledges the Spanish Ministry of Education for the FPU 15/03636 Ph.D. scholarship. SR acknowledges the Spanish Ministry of Economy for the FPI SEV-2015-0554-16-4 Ph.D. scholarship. The work of DRI is supported by the Spanish Ministry of Economy and Innovation program MTM2014-56949-C3-1-R,  and the AXA-ICMAT Chair on Adversarial Risk Analysis.

\bibliography{mybibfile}
\bibliographystyle{unsrt}

\end{document}